\documentclass{article}
\usepackage{emulateapj}

\lefthead{Nevalainen J., Markevitch M and  Forman W.}
\righthead{The Baryonic and Dark matter Distributions in Abell 401}

\begin{document}

\submitted{Accepted to ApJ}

\title{The Baryonic and Dark Matter Distributions in Abell 401}

\author{J. Nevalainen\altaffilmark{1}, M. Markevitch\altaffilmark{2} and W. Forman}
\affil{Harvard Smithsonian Center for Astrophysics, Cambridge, USA}

\altaffiltext{1}{Observatory, University of Helsinki, Finland}
\altaffiltext{2}{Space Research Institute, Russian Acad. of Sci.}


\begin{abstract}

We combine spatially resolved ASCA temperature data with ROSAT imaging data to constrain the total mass distribution in the cluster A401, assuming that the cluster is in hydrostatic equilibrium, but without 
the assumption of gas isothermality. We obtain a total mass within the X-ray core (290 $h_{50}^{-1}$ kpc) of
$1.2^{+0.1}_{-0.5} \times 10^{14} \, h_{50}^{-1} \, M_{\odot}$
at the 90 \% confidence level, 1.3 times larger than the isothermal estimate. 
The total mass within $r_{500}$ (1.7 $h_{50}^{-1}$ Mpc) is
$M_{500}$ = $0.9^{+0.3}_{-0.2} \times 10^{15} \, h_{50}^{-1} \, M_{\odot}$ 
at 90 \% confidence, in agreement with the optical virial mass estimate, and 1.2 times smaller than the isothermal estimate. Our $M_{500}$ value is 1.7 times smaller than that estimated using the 
mass-temperature scaling law predicted by simulations. The best fit dark matter density profile scales as $r^{-3.1}$ at large radii, which is consistent with the Navarro, Frenk \& White (NFW) ``universal 
profile'' as well as the King profile of the galaxy density in A401. From the imaging data, the gas density profile is shallower than the dark matter profile, scaling as $r^{-2.1}$ at large radii, leading to 
a monotonically increasing gas mass fraction with radius. Within $r_{500}$ the gas mass fraction reaches a value of 
$f_{gas} = 0.21^{+0.06}_{-0.05} \ h_{50}^{-3/2}$
(90 \% confidence errors). Assuming that $f_{gas}$ (plus an estimate of the stellar mass) is the universal value of the baryon fraction, we estimate the 90 \% confidence upper limit of the cosmological matter
density to be $\Omega_{m} < 0.31$, in conflict with an Einstein-deSitter universe.
Even though the NFW dark matter density profile is statistically consistent with the temperature data, its central temperature cusp would lead to convective instability at the center, because the gas 
density does not have a corresponding peak. One way to reconcile a cusp-shaped total mass profile with the observed gas density profile, regardless of the temperature data, is to introduce a significant 
non-thermal pressure in the center. Such a pressure must satisfy the hydrostatic equilibrium condition without inducing turbulence. Alternately, significant mass drop-out from the cooling flow would make the
temperature less peaked, and the NFW profile acceptable. However, the quality of data is not adequate to test this possibility.

\end{abstract}


\section{Introduction}

Determining mass components of clusters of galaxies is an important task in observational cosmology, since clusters form through the collapse of a large volume of primordial matter, and as such may provide a
representative sample of the universe as a whole (e.g. \cite{wnef}). Assuming hydrostatic equilibrium, the total mass of a cluster can be determined from the intracluster gas temperature and density 
distributions (\cite{bahc}, \cite{math}). Until recently, most hydrostatic X-ray mass estimates have been made  assuming that the gas is isothermal at the average broad beam temperature. ASCA observations 
provide spatially resolved X-ray spectroscopic measurements of hot clusters, and yield the 2D temperature structure of clusters. Indeed, a large number of ASCA clusters shows that the temperature declines with
increasing radius from the center (\cite{mar2}), in qualitative accordance with hydrodynamic cluster simulations (e.g. \cite{evr96}, \cite{bn}). This implies that the real hydrostatic mass at large cluster 
radii is smaller than that derived assuming isothermality. Consequently, the gas mass fraction is larger and the ``baryon catastrophe'' even more pronounced, compared to isothermal estimates 
(e.g. White \& Fabian 1995). In this paper, we estimate the total mass for the A401 cluster, using the actual temperature profile.
Our method is essentially the one used for A2163 (\cite{mar96}) and A2256 (\cite{mv2}). We assume that the cluster is in hydrostatic equilibrium and model the dark matter component using several different 
functional forms. We use the ROSAT data to fix the gas density profile and fit hydrostatic temperature models, as a function of dark matter density parameters, to the ASCA data.

A401 (z = 0.0748) is suitable for measuring the dark and total mass distributions, since it is fairly bright and hot ($\sim$ 8 keV) allowing accurate temperature determinations with ASCA. 
Also, the ASCA field of view covers the cluster to $r_{500}$ (the radius where the mean interior density equals 500 times the critical density, approximately the radius inside which hydrostatic equilibrium 
holds). A401 also has been observed with the ROSAT PSPC, which gives an accurate estimate of the gas content of A401 and shows no obvious substructure and no significant deviations from azimuthal symmetry. 
The ASCA 2D temperature map (\cite{mar2}) shows no strong asymmetric variation. Thus the assumption of hydrostatic equilibrium is likely to be valid. 
A401 is peculiar in the sense that it shows no significant evidence of a cooling flow (see Peres et al. 1998, who find only an upper limit for the mass deposition rate), even though it has a prominent cD 
galaxy at its center and no evidence of recent merger activity either in the temperature map or in the X-ray image.
However, the ROSAT HRI image shows a linear structure in the neighboring cluster A399, pointing towards A401, as possible evidence of some past interaction (\cite{fab}). The lack of a significant cooling flow 
simplifies our analysis since we need not consider multicomponent temperature models in the center. We use $H_{0} \equiv 50 \, h_{50} \, {\rm km} \, {\rm s}^{-1} \, {\rm Mpc}^{-1}$, $\Omega = 1$ and report 
90\% confidence intervals throughout the paper.

\section{ROSAT ANALYSIS}

\subsection{Data reduction}
\label{redu}
The ROSAT data consist of two PSPC pointings of A401, rp800235n00 and rp800182n00, the former taken on July 29, 1992 and the latter on January 22, 1992. The total exposure times are
7457 s and 6735 s respectively. Reductions were carried out using Snowden's Soft X-Ray Background programs (\cite{sno}), which reduced the total exposure to 11.7 ks. 
The spatial analysis was restricted to the band of 0.44 - 2.04 keV (Snowden's bands R4-R7) to improve sensitivity over the X-ray background (see \cite{dav}). The non-cosmic X-ray background was subtracted 
using Snowden's code. All detector and telescope effects, including vignetting, the mirror support structure shadows and varying detector quantum efficiency, together with the aspect and livetime information 
are incorporated in exposure maps  for each band. All analysis was done in each of the four bands, for both pointings, the resulting images were divided by the corresponding exposure maps and 
combined. The surface brightness contour map (smoothed by a Gaussian with $\sigma$ = 1$'$) is shown in Figure 1. The data show no obvious substructure and no strong deviations from azimuthal symmetry, implying 
that the assumption of hydrostatic equilibrium is likely to be valid.

\subsection{Spatial analysis}
\label{spat}
With our spatial analysis, we wish to address two questions. First, how far from the cluster center can we significantly detect the cluster emission, and second, how is the cluster gas distributed.
Figure \ref{fig1} shows that the surface brightness (cluster + background) reaches a constant level at a distance of about 30$'$ from the center (= 3.4 $h_{50}^{-1} \, Mpc$) 
so we measure the Cosmic X-Ray Background, plus any residual detector background, from the same image at radii between 30$'$
and 52$'$. To exclude any contribution from the nearby cluster A399 we excluded an azimuthal sector centered at A401, directed at A399 (154$^{\circ}$ clockwise from the north),
from our analysis. To determine the region to be excluded, we increased the angular extent of the excluded sector until the background level reached a minimum value, with a sector width of
160$^{\circ}$. Thus, the contribution of A399 to the 30$'$ - 52$'$ annulus (and inner annuli as well) becomes negligible outside this sector. 
We use the minimum value estimated above, 1.8 $\times 10^{-4}$ counts s$^{-1}$ arcmin$^{-2}$, as the total background value.
  
We excluded point sources and background fluctuations from our analysis. Around a radius of $20'$ ($18'$ - $27'$) it is not clear whether the small fluctuations are associated with cluster emission or not.
The detector support rib near 20$'$ may be causing some residual effects. This led us to two schemes, where we masked out, in addition to unambiguous point sources at other radii, i) all point-like sources 
between $18'$ - $27'$ (a conservative scheme) and ii) none of those sources (a non-conservative scheme), the truth being somewhere between. 

We generated a radial surface brightness profile in concentric annuli ranging from $15''$ at the center to $10'$ at a radial distance of $50'$. The signal-to-noise ratios as a function of radius include 5\% of 
the background value as a systematic error added in quadrature. In the conservative scheme, we can detect the cluster gas emission with 2.6 $\sigma$ significance at $r = 25'$, and in the non-conservative scheme 
with 3.8 $\sigma$ significance at the same radius, so we conclude that we detect the cluster gas at $ \sim 3 \sigma$ up to $25'$ (2.9 \  h$_{50}^{-1}$ Mpc). In our further analysis we consider only the profile 
of the conservative scheme.

We fitted the observed profile with the
$\beta$- model plus background
 
\begin{equation}
I(b) = I_{0}\left(1 + \left(\frac{b}{a_{x}}\right)^{2}\right)^{(-3 \beta + \frac{1}{2})}
\label{cav} + BGD
\end{equation}
(\cite{cava}), where $b$ is the projected radius and  
background BGD is fixed to the value found above. We used XSPEC to convolve the surface brightness model through a spatial response matrix (constructed from the ROSAT PSF at 1 keV, for a spherically symmetric 
on-axis source) and to compare the convolved profile with the data.  
The brightness in the two innermost bins ($r < 30''$) rises 2.1 $\sigma$ and 1.2 $\sigma$, or 47\% and 12\%, above the best fit $\beta$ - model, respectively.
This behaviour is consistent with Peres et al. (1998) who found that the A401 data, consistent with no cooling flow, did allow a mass accretion rate up to 120 $M_{\odot}$/yr
in the center of A401 within a cooling radius of $0.7^{+0.6}_{-0.7}$ arcmin. Even though the central excess in our data is statistically not very significant, we excluded our two innermost bins from the 
$\beta$ fit to prevent any bias towards small values of the core radius. 
We find an acceptable fit in the radial range 0.5$'$-52$'$ (see Figure
\ref{surfbright_plot} and Table \ref{tab1}), with best
fit parameters and  90 \% confidence errors of 
$a_{x} = 2.56 \pm 0.14$ arcmin (= $294 \pm 16 \ h_{50}^{-1}$ kpc),
$\beta = 0.70 \pm 0.02 $,
I$_{0} = 5.6 \pm 0.3 \ \times (10^{-2}$ counts s$^{-1}$ arcmin$^{-2}$) at $r = 0'$,
with $\chi^2$ = 49.7 for 55 degrees of freedom. The confidence contours for $a_{x}$ and $\beta$ are shown in Figure \ref{confcont_plot}.
Our values of $a_{x}$ and $\beta$ are consistent with the results of another study of the ROSAT data of A401 (Vikhlinin et al. 1998).

If we assume that the intracluster gas is isothermal and spherically symmetric, the best-fit parameters $a_{x}$ and $\beta$ will determine the shape of the gas density profile by the equation:
\begin{equation}
\rho_{gas}(r) = \rho_{gas}(0)\left(1 + \left(\frac{r}{a_{x}}\right)^{2}\right)^{ -\frac{3}{2} \beta}
\end{equation}
A401 is hot and even the temperature variations such as those detected by ASCA do not significantly affect the brightness in the ROSAT band.

We obtained the normalization (as in \cite{vikh}) $\rho_{gas}(0) = 1.6 \times 10^{14}$M$_{\odot}$ Mpc$^{-3}$, or 1.1 $\times 10^{-26}$ g cm$^{-3}$, by equating the emission measure calculated from the above 
equation, with an observed value of 16.7 $\times 10^{67}$ cm$^{-3}$ inside a cylinder with $r = 2$ $h_{50}^{-1}$ Mpc radius, centered at the cluster brightness peak. 

The gas mass in this best fit model inside the sphere of $r_{500}$ (=$15.1'$ = 1.7 $h_{50}^{-1}$ Mpc, calculated in Section \ref{coreres}) is
\begin{equation}
 M_{gas}(\le r_{500})  = 2.01 \pm 0.08  \times 10^{14} \, h_{50}^{-5/2} \, M_{\odot}
\label{gasmass}
\end{equation}
White \& Fabian (1995) give a gas mass value of $ 1.32 \pm 0.07 \times  10^{14}$ \, M$_{\odot}$ inside 1.3 $h_{50}^{-1}$ Mpc  whereas our corresponding value at the same radius is 
$1.4 \times  10^{14} \, M_{\odot}$, consistent with theirs. The error interval in (\ref{gasmass}) corresponds to the 90\% confidence region in $a_{x}$ - $\beta$ - space in our global fit 
(see Figure \ref{confcont_plot}). The gas mass error within $r_{500}$ is negligible with respect to the other errors in the quantities we are interested in this work. Therefore we ignore the above error 
in our further analysis.

\section{TEMPERATURE DATA}

\subsection{ASCA} 
\label{ascadata}
The gas temperature distribution is obtained from the ASCA spectral data of A401, excluding a sector towards A399, as described in detail in Markevitch et al. (1998).
The data are divided into four concentric radial bins, 0$'$-2$'$-5$'$-9$'$-16$'$ (0-0.23-0.57-1.0-1.8 h$_{50}^{-1}$ Mpc). 
The temperature errors were determined by generating Monte - Carlo data sets which properly  account for the statistical and systematic uncertainties (including PSF effects). 

Figure \ref{temperatures_plot} shows the best fit projected temperatures and 1$\sigma$ errors in each of the four radial bins.
We note a slight (by a factor of 1.2 or 1 $\sigma$) increase  of the temperature in the center, compared to a quite constant value in bins 2$'$-5$'$-9$'$.
The temperature in the 9$'$-16$'$ bin falls by a factor of 1.8 or 2.7 $\sigma$ below the value in the 5$'$-9$'$ annulus. Similar radial temperature declines have
been observed in a large sample of ASCA clusters (\cite{mar2}). The single temperature fit gives a mean temperature of
kT = $ 8.0 \pm 0.4 $ keV (\cite{mar2}), which is consistent with the EINSTEIN MPC value
kT = $7.8^{+1.1}_{-0.9}$ keV (\cite{dav2}).
The temperature profile values are consistent with the single temperature only by a probability of smaller than $10^{-6}$, therefore A401 is significantly non-isothermal.

We note that even though at the cluster center we saw a slight brightness excess ($\sim 30$ \%) compared to the $\beta$-model, the sky area covered by the central bins 
($r < 0.5'$) is only 6\% of the area covered by the innermost ASCA temperature bin of $r <2'$. Therefore the contribution of the brightness excess to the total emission from the central temperature bin is
negligible, and we ignore the effect of a possibly different temperature for this excess emission.

The derivation of cluster temperature profiles from the ASCA data is not straightforward due to the wide and energy-dependent PSF. Takahashi et al. (1995) showed that if the PSF effect is neglected, an 
intrinsically isothermal cluster will appear significantly hotter with ASCA at large radius (a 7 keV cluster would appear to have a temperature of $\sim$ 20 keV at a 20$'$ radius). For A401, Fujita et 
al.\ (1996) analyzed the same ASCA dataset and derived an approximately constant temperature profile. From the description in their paper, it appears that they did not properly include PSF scattering effects. 
This would have an effect of diminishing the radial temperature decline, consistent with the difference of the two results. ROSAT PSPC data on A401 in the 0.2--2 keV band were also analyzed by Irwin et al. 
(1999), who derive a temperature increase with radius. Taking into account the PSPC calibration uncertainty which is very significant for determining temperatures of hot clusters with PSPC (see, e.g., Markevitch
\& Vikhlinin 1997a), their results would probably be consistent with ours. Among other clusters in the Markevitch et al.\ (1998) sample, ASCA data for A4059 and MKW3s were recently independently analyzed by 
Kikuchi et al.\ (1999) using an independent method. Although these authors do not find radial temperature gradients as strong as in Markevitch et al. (1998), they also do not detect the strong cooling flows 
that are known to exist in those clusters (e.g., Peres et al.\ 1998), which indicates possible problems with their results. For another cluster from that sample, Hydra-A, the Markevitch et al. \ (1998)  
declining temperature profile is in good agreement with an analysis by Ikebe et al.\ (1996) using a different technique. A strong radial temperature decline was also recently derived for A2218 by 
Cannon, Ponman, \& Hobbs (1999) using yet another method. Still another technique developed by Churazov et al. (1996) yields a temperature structure in Coma and A1367 consistent with the Markevitch et al.\ 
(1998) method (see \cite{donn1}, \cite{donn2}). Finally, the A401 temperature profile that we use in this work, is similar to profiles of a large sample of ASCA clusters (\cite{mar2}), when compared in the 
physically meaningful units of the radius of a given overdensity. Therefore it appears unlikely that the observed temperature decline is caused by any unknown detector-dependent systematic effect. 

\subsection{ROSAT} 
\label{rosattemp}
In addition to our four ASCA temperature points, we have some crude temperature information from the ROSAT data. Since we have a significant ($\sim$ 3 $\sigma$) detection of the cluster gas in the 
22$'$-25$'$ (= 2.5-2.9 $h_{50}^{-1}$ Mpc) annulus we know that the gas temperature there must exceed zero. Therefore we introduce a ROSAT temperature point kT $>$ 0 keV at 22$'$ - 25$'$.

\section{MASS CALCULATION}

\label{masscalc}
We will now use the ASCA and ROSAT temperature data to estimate the total mass of A401. For this, we assume that A401 is spherically symmetric and that its gas is in hydrostatic equilibrium (as indicated in 
Section \ref{redu}). From this condition the total mass within a sphere of radius $r$ can be written as (e.g. \cite{sar}):

\begin{equation}
M_{tot}(\le r) = 3.70 \times10^{14} M_{\odot} {T(r) \over {\rm 10 keV}} {r \over {\rm Mpc}} \left( - {{d \ln{\rho_{gas}}} \over {d \ln{r}}} - {{d \ln{T}} \over {d \ln{r}}} \right),
\label{hydreq}
\end{equation}
using $\mu = 0.60$.

We consider the total mass consisting of stellar mass in galaxies, intracluster gas, and dark matter. We estimate the amount of stellar mass in galaxies using Dressler's (1978) King profile fit to the galaxy 
distribution in A401 with a core radius of 0.4 $h_{50}^{-1}$ Mpc and a V band luminosity of $1.5 \times 10^{12}$ $h_{50}^{-2}$ L$_{\odot}$ inside 1 core radius. Even though the fit extends only to a 12$'$
radius, we extrapolate this distribution to 15.1$'$  (= $r_{500}$) using the above King profile, and obtain a V band luminosity of $1.0 \times 10^{13}$ $h_{50}^{-2}$ L$_{\odot}$. We convert this value to galaxy
mass, using a mass-to-light ratio of 3.2 M$_{\odot}$/L$_{\odot}$ $h_{50}$ (taken from \cite{wnef}), assuming a Coma-like luminosity function for galaxies in A401, and an M/L relation from van der Marel (1991).
We find a  stellar mass in galaxies of 3.3 $ \times 10^{13}$ $h_{50}^{-1}$ M$_{\odot}$, or 16  $h_{50}^{-3/2}$ \% of our gas mass value, or 4 \% of our total mass value within $r_{500}$. Since this is much less
than the uncertainty of our total mass estimates, we will not model this component separately, but rather include it in the dark matter model. This does not introduce any ambiguity in the interpretation of dark
matter parameters (except for the normalization), since our best fit dark matter profile scales like the King profile at large radii (see below) and  the core radius of this model is similar to the best fit
King profile of the galaxy distribution in A401 (Dressler 1978). Quantitatively, the total mass is given as
\begin{equation}
M_{tot}(\le r) = \int_{0}^{r} 4 \pi r^{2}(\rho_{dark} + \rho_{gas}) \,dr ,
\end{equation}
where $\rho_{dark}$ and $\rho_{gas}$ denote the density profiles of the dark matter plus stars and gas, respectively.  

In principle, the mass can be calculated directly from Equation \ref{hydreq}, if the gas temperature profile is known in detail. However, our temperature profile is not of sufficient quality  to allow this 
procedure. We therefore use an indirect method which assumes various models for the total mass radial distribution, calculates the corresponding temperature profiles, and compares them to the data, looking for 
acceptable models. Following Markevitch \& Vikhlinin (1997b), we model the dark matter density distribution $\rho_{d}$ using two different functional forms, which together approximate a wide range of physically
motivated spherically symmetric distributions. A constant core model has a dark matter density given by
\begin{equation}
\rho_{dark} \propto \left(1 + \frac{r^2}{a_d^2}\right)^{- \alpha/2},
\label{coremodel}
\end{equation}
and the central cusp profile is described as:
\begin{equation}
\rho_{dark} \propto \left(\frac{r}{a_{d}}\right)^{- \eta} \left(1 + \frac{r}{a_{d}}\right)^{\eta - \alpha}.
\label{cuspmodel}
\end{equation}
In the equations above, $a_{d}$ is the scale length of the dark matter distribution. In the cusp model (Eq. \ref{cuspmodel}), the first term describes the cusp behaviour near the center. In both models, the 
density at large radii scales as $r^{-\alpha}$. With $\eta$ = 1 and $\alpha$ = 3, the cusp model corresponds to the ``universal density profile'' which Navarro et al. (1995, 1997, NFW hereafter) show to be a 
good description of cluster CDM halos in simulations of hierarchical clustering. Since the quality of the data is not adequate for deriving the values of all the parameters independently, we fix  $\eta$ = 1 in 
the cusp models, as suggested by the NFW simulations, but vary the other parameters.

As in Markevitch \& Vikhlinin (1997b), we solve the hydrostatic equilibrium equation (Eq. \ref{hydreq}) for the gas temperature as a function of radius and gas and dark matter density parameters 
(using the $\beta$ model for the gas density profile):

\begin{equation}
T(r) = {(1 + x^{2})^{3\beta/2}} \left[T_{0} - A \int_{0}^{x} {(1 +
y^{2})^{-3\beta/2}} \ {I(y) \over y^{2}}  d y \right],
\label{A5}
\end{equation}
where
$ x \equiv r/a_{x}$,
\begin{equation}
A \equiv {{4 \pi} \over {3.70 \times 10^{13}}} {{\left({{a_{x}} \over {\rm Mpc}} \right)}{^2}} {{\rho_{gas}(0)} \over {M_{\odot} \ {\rm Mpc}^{-3}}}  {\rm keV}
\end{equation}
and
\begin{equation}
I(y) \equiv \int_{0}^{y} z^{2} (1 + z^{2})^{-3\beta/2} \ d z + {\rho_{d1} \over \rho_{gas}(0)} \int_{0}^{y} f_{d}(z) d z,
\label{A4}
\end{equation}
where $f_{d} \equiv \rho_{dark}(x) / \rho_{d1}$, and  $\rho_{d1}$ is the dark matter density at the X-ray core radius $a_{x}$, and $\rho_{dark}(x)$ is given by either Equation (\ref{coremodel}) or 
(\ref{cuspmodel}). $T_{0}$ and $\rho_{gas}(0)$ are the gas temperature and density at $x = 0$ . 

For the gas density distribution, we use the $\beta$ model parameters derived from the ROSAT surface brightness analysis (Section \ref{spat}). The remaining parameters to be fitted to the temperature
data are $\alpha$, $a_{d}$, $\rho_{d1}$ and $T_{0}$.

While computing the temperature profiles, we use analytic solutions for the integrals in Equation (\ref{A4}) in the cluster center to avoid the numeric effects of the singularity in the integral in Equation 
(\ref{A5}). Beyond 0.2 $a_{x}$ we switch to numerical integration, preserving continuity. Step sizes for the numerical integrations are chosen to achieve a 1\% accuracy in the computed temperatures. 

For each set of dark matter profile parameters and $T_{0}$ we computed the 3D temperature profile, and projected it to the observed 2D ASCA annuli, weighting the line of sight temperatures with the emission 
measure of each volume element intersected by the ASCA annuli. The projected model temperatures were then compared with the measured ASCA temperature values. The parameter values were changed
iteratively to minimize $\chi^{2}$.
 
To incorporate the ROSAT temperature point (Section \ref{rosattemp}) in the fitting procedure, we do the following:
if the model temperature at 22$'$ - 25$'$  becomes negative, an exponential increment is added to $\chi^2$ (the more negative the value, the higher the increment), but as soon as the trial value in the fit 
becomes positive, the contribution of that data point to $\chi^2$ vanishes. This arrangement will ``steer'' the fitting procedure smoothly towards positive temperature values in the 22$'$-25$'$ bin, without 
bias towards any arbitrary temperature value.

The model temperature profile is very sensitive to small changes in the parameter values. On the other hand, the temperature errors and the widths of the radial bins are quite large, due to poor statistics.
Therefore, a large range of parameter values gives a good description of the data. These two features  led to difficulties using standard minimization routines, which often find local $\chi^2$-minima.
We dealt with this problem by applying the simulated annealing method
(\cite{press}).

\section{RESULTS}

\label{thermstab}
Before presenting the results of our mass fits, we briefly address the question of convective  stability, since some of our model temperature profiles, formally allowed by the data,  have strong 
gradients at large radii and at the very center (especially the cusp models). These strong temperature gradients may not be consistent with the requirement of convective stability.
If the gas at radius $r$ is to be stable against convection, the effective polytropic index at that radius, defined as
\begin{equation}
\gamma(r) = {{d\log T(r)} \over {d\log \rho(r)}} + 1,
\label{gamma}
\end{equation}
must be less than $\frac{5}{3}$. During a dynamical time, all convective instabilities should have been erased, and clusters in general should be convectively stable, therefore strong temperature gradients 
cannot exist. However, at large radii, hydrostatic equilibrium may not have time to establish. Simulations (\cite{evr96}) suggest that $r_{500}$ (the radius where the mean interior density is 500 times the 
critical density) provides a conservative upper limit for the radius inside which the gas should be hydrostatic. However, in many simulated clusters, hydrostatic balance holds at evenlarger radii. 
For A401, $r_{500} = 15.1'$, and $r_{150} = 25'$, the radius where the mean interior density is 150 times the critical density. Hence, at the radial range of our ASCA data ($r < 16'$) the hydrostatic
equilibrium assumption is likely to be valid, and we are justified to require that the model temperature profiles be convectively stable. At the maximum radius of ROSAT detection (25$'$) the case is less 
certain. In the Evrard et al. (1996) simulations, some clusters exhibit significant gas bulk motion at these radii. Depending on the cosmological scenario, at $r_{150}$ the kinetic pressure may be comparable 
to the thermal pressure, and the systems may be far from hydrostatic equilibrium. Since A401 is an apparently relaxed cluster, strong deviations from equilibrium are unlikely even at that radius.
In order to be conservative, we will not apply the above convective stability constraint, $\gamma \le \frac{5}{3}$, at radii beyond $r_{500}$. Also, we do not apply it to our solutions 
at $r < 0.5'$ because at that radial range the cluster brightness does not agree with the $\beta$ model.

While using the ROSAT temperature information at $25'$ in our fits, we are extrapolating the hydrostatic temperature profile to a possibly non-hydrostatic region of the cluster (out to $r_{150}$), but this 
should not introduce a significant error. Cluster formation simulations show that temperature profiles of clusters decline with the radius without any dramatic change in the temperature profiles between 
$r_{500}$ and $r_{150}$, so that our extrapolation is justified, even though the kinetic pressure may become comparable to the thermal pressure at $r_{150}$. Furthermore, if the gravity of the total mass 
inside r$_{150}$ balances the thermal plus kinetic pressure at that radius, for a given mass distribution the temperature implied by the hydrostatic model is {\it higher} than the observed one. 
Hence, in our fitting the requirement that the hydrostatic model temperature exceed zero at r$_{150}$, is conservative. A possible complication, a T$_{e}$ - T$_{i}$ nonequality at large radii 
(e.g. \cite{fox}, \cite{ett}, \cite{tak}) would have the same effect.

\vspace{1cm}

\subsection{Core model}
\label{coreres}
We now discuss results of fits using the dark mass profile of the core model (Eq. \ref{coremodel}). To avoid local minima, we adopted a scheme to fix $\alpha$ at several values over an interesting range
and to fit the other parameters. This was necessary, since the effect of this parameter is most significant at large radii where we do not have ASCA temperature data. The best fit core model is acceptable, 
with $\chi^2$ = 1.29 for 4 parameters and 5 data points (see Figures \ref{temperatures_plot} and \ref{masses_plot} and Table \ref{tab1}). The best fit core model gives a value for the dark mass 
M$_{d}$ = 0.73 $\times 10^{15} \ h_{50}^{-1}$ M$_{\odot}$
within $r < r_{500}$, with 
T$_{0}$ = 10.1 keV,
$\rho_{d1}$ = 4.3 $\times 10^{-26}$ g cm$^{-3}$ 
(which corresponds to the central dark matter density of
1.5 $\times 10^{-25}$ g cm$^{-3}$),
$a_{d}$ = 2.29 arcmin (= 260 \ $h_{50}^{-1}$ kpc),
$\alpha$ = 3.1.
The central dark matter density is 14 times that of the gas. The core radii of gas and dark matter models are quite similar, and the dark matter density at large radii falls faster, scaling as r$^{-3.1}$, 
whereas the gas density falls as r$^{-2.1}$. To check whether the solution agrees with the requirement of convective stability, we computed the effective polytropic index as a function of radius, using Equation
\ref{gamma} and used the gas distribution from the best fit $\beta$ model. This best-fit model is convectively stable in the $ r = 0.5' - 15.1'$ range. The overdensity in this best-fit model drops below 500 at
radius
\begin{equation}
r_{500} = 15.1' = 1.7 \ h_{50}^{-1} \ {\rm Mpc}.
\label{r500eq}
\end{equation}

To propagate the errors of the temperature profile data to our mass values, we fit a large number of Monte - Carlo temperature profiles with added random errors (see Section \ref{ascadata}) using the same 
approach as for the best-fit temperature profile. We noticed that most models had a tendency for the temperature to fall to zero at radii below $25'$ (because of the strong decline of the temperature between 
the two outermost ASCA bins), therefore our ROSAT temperature information at large radius provides a powerful constraint. We rejected unphysical models which gave infinite temperatures at large radii. 
Each Monte - Carlo model was checked against the convective stability constraint, and rejected if $\gamma > \frac{5}{3}$ in the $r = 0.5' - 15.1'$ range.
The excluded models are those with the sharpest temperature peaks in the center, or those with the smallest core radii, which are also the models with smallest $\alpha$ due to the parameter correlations. 
The mass in the models with the smallest $\alpha$ values increases fastest with radius, therefore large masses are constrained more strongly. From the distribution of the accepted Monte - Carlo models,
we determine the 1 $\sigma$ scatter of the mass values as a function of the radius. We convert these values to 90\% confidence values, assuming a Gaussian probability distribution. Although in general we cannot
constrain the dark matter model parameters independently, some of these parameters are correlated and the corresponding mass values vary within a relatively narrow range and can be reasonably constrained.

Since the total mass is proportional to the local logarithmic derivative of the gas density distribution ($\propto \beta$ in Eq. (2)), we quantified the effects of the uncertainty of the local $\beta$ value 
and its possible deviations from the global value (see Section \ref{spat}) that we used in our mass computation. We divided the ROSAT profile into radial ranges r = 0.5-5$'$, 5-10$'$, 10-20$'$, 15-30$'$ and 
fit these profiles with the $\beta$ model, fixing the core radius to its global value. Within 15$'$ the $\beta$ values are similar to the global value, while beyond 15$'$ the profile becomes slightly (but not 
significantly) steeper. We added the local $\beta$ uncertainty in quadrature to the total mass uncertainty, which gave a very small effect within $r_{500}$. Our final mass values for the core model 
at the gas core radius $r = a_x$ (= 2.6$'$ = 290 $h_{50}^{-1}$ kpc) and at $r_{500}$ (= 15.1$'$ = 1.7 Mpc) are 
\begin{equation}
M_{tot}(\le a_x) = 1.20^{+0.11}_{-0.49} \times 10^{14} h_{50}^{-1} \ M_{\odot}
\end{equation}
and
\begin{equation}
M_{tot}(\le r_{500}) = 0.94^{+0.24}_{-0.22} \times 10^{15} h_{50}^{-1} \ M_{\odot}.
\label{M500_eq}
\end{equation}
Figure \ref{masses_plot} shows the resulting mass profile and the corresponding f$_{gas}$ profile.

\subsection{Cusp model}
We now apply the cusp model given in Eq. \ref{cuspmodel}. The best fit cusp model ($\eta \equiv 1 $) has an unreasonably high value of the slope parameter $\alpha$, due to parameter correlations. However, for 
the best fit models, the total mass within $r_{500}$ does not depend significantly on $\alpha$. The best fit total mass within $r_{500}$ is 0.96  $\times 10^{15} \, h_{50}^{-1} \, M_{\odot}$,  almost identical 
to the best fit value for the core model obtained above (Eq. \ref{M500_eq}). In the radial range $a_x < r < r_{500}$ the enclosed masses in the best fit core and cusp models differ by not more than 5\%, a 
deviation that is negligible compared to the mass uncertainties. The scatter of mass values for the cusp model is smaller than that for the core model, so that at each radius the 1 $\sigma$ interval of the cusp
model masses lies within that of the core model. Therefore, the mass values and errors obtained earlier with the core model will be our final values (see Table \ref{tab2}).

Even though the cusp model gives an acceptable fit to the temperature data, because of the centrally peaked form of this model it always violates the convective stability constraint at the center, even at
radii $r > 0.5'$ where the gas profile is well defined by the $\beta$ model. Models with $\eta < 1$ have a less prominent peak, but as $\eta$ approaches zero, the models essentially approach the constant 
core model already considered in Section \ref{coreres}. The masses in the best fit models with different $\eta$ values within $r_{500}$ equal that of the $\eta \equiv 1$ model. Since the gas distribution in
A401 is well represented by a $\beta$ model inwards to a rather small radius ($0.5'$), one way to reconcile the cusp model, such as those predicted by simulations, with the observed gas density profile is to 
introduce a significant non-thermal pressure in the center (\cite{loeb}, \cite{mv2}). Such a pressure must satisfy the hydrostatic equilibrium condition without inducing turbulence, which would require, for
example, an equation of state $p_{\rm therm}+p_{\rm nontherm}\propto \rho^\gamma$ with $\gamma\ll 5/3$ (e.g., Landau \& Lifshitz 1959).

Note that A401 is rather unusual in that it has a cD galaxy but no significantly detected cooling flow and gas density peak usually found in cD clusters. 
However, the results of Peres et al. (1998) allow an upper limit of 120 $M_{\odot}/yr$ for the mass flow rate in A401. 
Significant mass deposition from the cooling flow would make the average gas temperature less peaked, and possibly the NFW profile acceptable. 
Unfortunately, the quality of the current data is not adequate to construct a proper two-phase modelling of the cluster medium to test this possibility.

\subsection{Comparing mass values}
\label{isot}
For comparison, we have calculated the mass profile under the traditional assumption of isothermality. The emission-weighted temperature model of the gas excluding the contaminating components for A401 gives 
T$_{X} = 8.0 \pm 0.4$  
(\cite{mar2}).
Assuming a constant temperature, Equation \ref{hydreq} reduces to 
\begin{equation}
M_{tot}(\le r) = 1.11 \times 10^{15} \beta {{T_{x}} \over {10 \ {\rm keV}}} {{r} \over {\rm Mpc}}{{(r/a_{x})^2} \over {1 + (r/a_{x})^2 }} M_{\odot}.
\label{isot_eq}
\end{equation}
Figure \ref{masses_plot}b shows this mass profile together with that derived using the observed temperature profile. At $r = a_{x}$ (= 2.6$' = 0.29 \, h_{50}^{-1}$ Mpc) the mass derived using the observed 
temperature profile exceeds the isothermal mass by a factor of 1.3, both agree at a radius of 11$' = 1.2 \, h_{50}^{-1}$ Mpc and at 15.1$'$ (= 1.7 Mpc = $r_{500}$) our value falls to 0.9 of the isothermal 
value. Qualitatively similar behaviour was found by Markevitch \& Vikhlinin (1997) for A2256, where the effect had a larger magnitude due to the somewhat steeper temperature decline and wider radial coverage.

Using the deprojection technique, and the broad beam EINSTEIN MPC temperature of kT = $7.8 \pm 0.6$ keV (\cite{dav2}), White \& Fabian (1995) and White et al. (1997)  obtained gravitating mass values of
1.01, 0.59 and 1.07  $ \times 10^{15} \, h_{50}^{-1} M_{\odot}$ 
inside radii of 
1.265, 0.862 and 1.380 h$_{50}^{-1}$ Mpc, 
respectively. These masses are higher than the isothermal masses obtained by evaluating Equation (\ref{isot_eq}). Our temperature profile masses at the same radii are
$0.74^{+0.17}_{-0.08}$, $0.53^{+0.13}_{-0.04}$ and $0.79^{+0.19}_{-0.11}  \times 10^{15} \ h_{50}^{-1}$, 
respectively, which are lower by factors of 1.4, 1.1 and 1.3 (2.6, 0.7 and 2.5 $\sigma$). This difference is in line with our finding above that, with increasing radius, the isothermal model overestimates 
the mass.

It is useful to compare our measurement with the cosmological simulations of Evrard et al. (1996), who obtained  cluster mass-temperature and radius-temperature scaling
laws of the form:
\begin{equation}
r_{500}(T_{X}) = 2.48 \pm 0.28 \ (T_{X}/10 keV)^{1/2} \ Mpc
\end{equation}
and
\begin{equation}
M_{500}(T_{X}) = 2.22 \pm 0.55 \ (T_{X}/10 keV)^{3/2} \ \times 10^{15} \ M_{\odot}.
\end{equation}
For a cluster with T$_{X} = 8.0 \pm 0.4$ keV, Evrard et al. (1996) predict
$r_{500} = (2.22 \pm 0.16)$   Mpc and
$M_{500} = (1.59 \pm 0.25)  \times 10^{15} \, M_{\odot}$.
Using our best fit core model, the corresponding values (Eq. \ref{r500eq} and  \ref{M500_eq}) are by factors of 1.3 and 1.7 (3.1 $\sigma$) lower than the values predicted by the scaling law, respectively.
In the case of A2256 (\cite{mv2}), a similar difference was found. The isothermal model (Equation \ref{isot_eq}) gives $M_{500} = 1.1 \times 10^{15} \ h_{50}^{-1}$, a factor 1.4 lower than the scaling law 
value. The difference is due to two effects: the simulated clusters have steeper gas density profiles and shallower temperature profiles than those observed.

Buote \& Canizares (1996) studied the ellipticity gradients of the ROSAT data for A401 and derived the total mass distributions in A401.
Their shape for the total mass density ($\rho_{tot} \propto r^{-4}$) is consistent with ours, but the normalization is very different.
Their total mass values within radii 0.8, 1.6, and 2.4 $h_{50}^{-1}$ Mpc are 1.47-1.70, 2.26-4.00 and 2.56-5.73  $\times 10^{15} h_{50}^{-1} M_{\odot}$, while our values at these radii
are 0.46-0.61, 0.71-1.10 and 0.73-1.51 $\times 10^{15} h_{50}^{-1} M_{\odot}$, smaller by factors of 2-4, 2-6 and 2-8, respectively. This behaviour for A401 is similar to what  Buote \& Canizares (1996) find
for A2199, for which they obtain 6 times larger total masses within 0.8  $h_{50}^{-1}$ Mpc, compared to a (isothermal) $\beta$ model estimate.

In the optical, the virial theorem (\cite{gir}) gives R$_{vir}$ = 4.6 h$_{50}^{-1}$ Mpc = 40.0 arcmin) and M$_{vir} = 2.74^{+0.92}_{-0.82}$ $ \times 10^{15} h_{50}^{-1} M_{\odot}$, while our values extrapolated
to that radius are
1.76 $^{+0.90}_{-1.15}  \times 10^{15} \  h_{50}^{-1}$ which is consistent
within 90\% confidence errors, but note that the extrapolated values are very uncertain.

\subsection{Baryonic fraction, $\Omega_{m}$}
The best fit models show that the dark matter density declines more rapidly than the gas density, which also means that the gas mass fraction $f_{gas}(<r) = M_{gas}(<r)/{M_{tot}(<r)}$ is a monotonically 
increasing function of radius (see Figure \ref{masses_plot}c). In A401, at $r_{500}$ the gas mass fraction reaches a value of
\begin{equation}
f_{gas}(\le r_{500}) = 0.21^{+0.06}_{-0.05} \, h_{50}^{-3/2}.
\label{fgaseq}
\end{equation}
This behaviour and value are similar to the results of a sample of EINSTEIN clusters (\cite{wf}), a sample of ROSAT clusters (\cite{dav}) and A2256 (\cite{mv2}). As shown by White et al. (1993),  $f_{gas}$ has 
important implication for the cosmological matter density parameter $\Omega_{m} = <\rho>/\rho_{crit}$. We define $\Upsilon$ as the ratio of the local baryon fraction $f_{b}$ in a cluster to the primordial 
value $\Omega_{b}
/ \Omega_{m}$. Therefore,
\begin{equation}
{{\Omega_{b}} \over  {\Omega_{m}}}  \Upsilon = f_{b}. 
\end{equation}
Assuming that the baryonic matter consists of the gas and stellar mass in the cluster, we write
\begin{equation}
f_{b} = {M_{gas} + M_{gal} \over M_{tot}} = f_{gas} + {M_{gal} \over M_{tot}}. 
\end{equation}
Hence,
\begin{equation}
\Omega_{m} = \Upsilon \, \Omega_{b}  {\left(f_{gas} + {M_{gal} \over M_{tot}}\right)}^{-1}
\label{omegaeq}
\end{equation}
We evaluate Equation \ref{omegaeq} at $r_{500}$ using
1) our $f_{gas}$ value (Equation \ref{fgaseq}),  
2) our galaxy mass estimate from Section \ref{masscalc},
3) a slight baryon diminution $\Upsilon(500) = 0.90$ suggested by simulations, (Frenk et al. 1996),
4) $\Omega_{b} h_{50}^{2} = 0.076 \pm 0.007$ (\cite{bur}),
and propagate the errors of $f_{gas}$ and $\Omega_{b}$.
Figure \ref{omega_plot} shows the resulting allowed parameter space of ($\Omega_{m}, H_{0}$).  
If some dark matter is baryonic, then $\Omega_{m}$ would decrease further.$\Omega_{m} = 1$ is allowed only by an unrealistically low value of the Hubble constant, $H_{0} < 8$ km s$^{-1}$ Mpc$^{-1}$. Using, 
for example a value $H_{0} = 68 \pm 8$ km s$^{-1}$ Mpc$^{-1}$ found from the analysis by Nevalainen \& Roos (1998) who studied the Cepheid metallicity effect on galaxy PL-relation distances calibrated
at LMC, we find a cosmologigal matter density parameter (90 \% confidence) of  
\begin{equation}
\Omega_{m} = 0.22^{+0.09}_{-0.08}, 
\end{equation}
which is consistent with the value obtained by combining all relevant current independent $\Omega_{m}$ estimates, $\Omega_{m} = 0.33 \pm 0.11$ (\cite{matts}).

\section{CONCLUSIONS}

Using spatially resolved ASCA spectroscopic data, we have constrained the dark matter distribution in A401, without the assumption of isothermality. The dark matter density in the best fit ``constant core'' 
model scales as r$^{-3.1}$ at large radii. Thus a well-known King model appears to describe the dark matter distribution, as well as the galaxy distribution in A401 (Dressler 1978). This slope is also the 
same as found by NWF in their simulations, although the simulated clusters also exhibit a central density cusp. 
For A401, such a cusp violates the convective stability condition in the cluster center, because the gas density is well described by a standard $\beta$ model. One way to reconcile a total mass profile 
having a cusp shape with the observed gas density profile is to introduce a significant non-thermal pressure in the center. Such a pressure must satisfy the hydrostatic equilibrium condition without inducing 
turbulence, which would require, for example, an equation of state $p_{\rm therm}+p_{\rm nontherm}\propto \rho^\gamma$ with $\gamma\ll 5/3$ (e.g., Landau \& Lifshitz 1959). Alternately, significant mass 
deposition from the cooling flow would make the temperature less peaked, and NFW profile acceptable. However, the quality of the data is not adequate to test this hypothesis.
Regardless of the presence or absence of a central cusp, the total mass within $r_{500}$ (1.7 $h_{50}^{-1}$ Mpc) is $0.94^{+0.24}_{-0.22} \times 10^{15} \, h_{50}^{-1} \, M_{\odot}$ at the 90 \% confidence. 
The mass within the X-ray core (290 $h_{50}^{-1}$ kpc) is a factor of 1.3 higher than the value from an isothermal analysis while at $r_{500}$ the value is 0.9 the isothermal value, which is qualitatively 
similar to the A2256 result (Markevitch \& Vikhlinin 1997). Our M$_{500}$ value is 1.7 times smaller than that predicted by the scaling law of Evrard et al. (1996). This discrepancy arises because
the simulations do not correctly predict the observed gas density and temperature profiles. The gas density profile is shallower than that of the dark matter, being proportional to  $r^{-2.1}$ at large radii. 
Hence the gas mass fraction increases with radius, with $f_{gas}(r_{500}) = 0.21^{+0.06}_{-0.05} \, h_{50}^{-3/2}$ (90 \% errors) at $r_{500}$. Assuming that the cluster matter content is representative
of that of the Universe, this implies $\Omega_{m} < 0.31$ at 90\% confidence, in conflict with the Einstein-deSitter Universe.

\acknowledgments
JN thanks Harvard Smithsonian Center for Astrophysics for the hospitality. JN thanks the Smithsonian Institute for a Predoctoral Fellowship, and the Finnish Academy for a supplementary grant. 
We are indebted to Dr A.Vikhlinin for several helpful discussions. We thank Prof. M.Roos and Dr A.Dressler for their help, and the referee for helpful comments.
WF and MM acknowledge support from NASA contract NAS8-39073.

\clearpage

\begin{deluxetable}{lr}
\tablecaption{Best-fit values with 90 \% confidence errors \tablenotemark{*} \label{tab1}}
\tablewidth{0pt}
\tablehead{
\colhead{Parameter} & \colhead{Value}}
\startdata
$a_x$ [arcmin]            & $2.56 \pm 0.14$ \\
$a_x$ [$h_{50}^{-1} \ {\rm kpc}$] & $294 \pm 16 $        \\
$\beta$ & $0.70 \pm {0.02}$ \\
I$_{0}$ [$10^{-2} \ {\rm c} \ {\rm s}^{-1} \ {\rm arcmin}^{-2}]$ & $5.6 \pm 0.3$ \\
$\chi^{2}$/d.o.f. & 49.7 / 55  \\
 &  \\
$\rho_{gas}$(0) [$10^{14}  \ h_{50}^{1/2} \ M_{\odot} \ {\rm Mpc}^{-3}$] & 1.6    \\
$\rho_{gas}$(0) [$10^{-26} \ h_{50}^{1/2} \ {\rm g} \ {\rm cm}^{-3}$] & 1.1   \\
&  \\
T$_{0}$ [keV]                          & 10.1 \\
$\rho_{d}$(0) [$10^{-25}$ g cm$^{-3}$] & 1.5  \\
$a_{d}$ [arcmin]                       & 2.29 \\
$a_{d}$ [$h_{50}^{-1} \ {\rm kpc}$]          & 260  \\
$\alpha$                               & 3.1  \\
\tablenotetext{*} {using $H_{0} \equiv 50 \ h_{50}$ \ km \ s$^{-1}$ \ Mpc$^{-1}$}
\enddata
\end{deluxetable}

\clearpage

\begin{deluxetable}{lcccc}
\tablecaption{Mass values with 90 \% confidence errors \tablenotemark{*} \label{tab2}}
\tablewidth{0pt}
\tablehead{\colhead{Radius} & \colhead{$M_{gal}$} & \colhead{$M_{gas}$} & \colhead{$M_{tot}$} & \colhead{$f_{gas}$ $\times$ $h_{50}^{3/2}$} \\
\colhead{[$h_{50}^{-1}$ Mpc]} & \colhead{[$ 10^{13} \
h_{50}^{-1} \ M_{\odot}$]} &\colhead{[$ 10^{14} \ h_{50}^{-1} \
M_{\odot}$]} & \colhead{[$ 10^{15} \ h_{50}^{-1} \ M_{\odot}$]} & \colhead{}} 
\startdata
0.29 (= $a_{x}$) & 0.23 &  0.11 & $0.120^{+0.011}_{-0.049}$ & $0.09^{+0.04}_{-0.01}$   \\
1.0                & 2.0  &  0.99 & $0.61^{+0.16}_{-0.05}$    & $0.16^{+0.01}_{-0.04}$   \\
1.7 (= $r_{500}$) & 3.3  &  2.01 & $0.94^{+0.24}_{-0.22}$    & $0.21^{+0.06}_{-0.05}$   \\
\tablenotetext{*} {using $H_{0} \equiv 50 \ h_{50}$ \ km \ s$^{-1}$ \ Mpc$^{-1}$}
\enddata
\end{deluxetable}

\clearpage

\figcaption{Contour map of the surface brightness of combined
rp800182n00 + rp800235n00 pointings, smoothed by a Gaussian with
$\sigma = 1'$.  The contour level values are 0.0005, 0.001, 0.002,
0.004, 0.008, 0.016 and 0.032 counts s$^{-1}$ arcmin$^{-2}$. A401 is
located in the center of the image, and the neighboring cluster A399 is located south-west from the image center. \label{fig1}}

\figcaption{ROSAT PSPC radial surface brightness profile together with
the PSF-convolved best-fit $\beta$ model.
The data with $r < 0.5'$ are not included in the fit.
Note the slight excess in the center.  \label{surfbright_plot}}

\figcaption{Confidence contours for the $a_{x}$ and $\beta$ parameters
of the surface brightness profile fit in Figure 2.  \label{confcont_plot}}

\figcaption{Crosses show projected ASCA temperatures with 1 $\sigma$
errors, and the lower limit from the ROSAT data. Thin solid lines show a representative set of temperature
models (before projection), allowed by the convective stability constraint (see text). Thick
solid line shows the best fit model (before projection). Thick dotted line shows the
values of this best model projected to the 2D ASCA bins, which are compared
with the ASCA data. \label{temperatures_plot}}

\figcaption{Mass distributions of the models plotted in Figure 4:  
(a) the dark matter and the gas densities, (b) the enclosed total masses with
the errors and the gas mass and 
(c) the gas mass fraction with errors. For comparison, values assuming isothermality (kT = 8.0 keV) are also shown.  Masses are evaluated using $H_{0} = 50$ km$^{-1}$ Mpc$^{-1}$ (total mass scales as  $H^{-1}$ and gas mass as  $H^{-5/2}$).\label{masses_plot}}

\figcaption{The 90 \% confidence area in parameter space of
[$H_{0},\Omega_{m}$], using the derived value of $f_{gas}$. A very low
value of $H_{0}$ is needed for $\Omega_{m}$= 1. The shaded area shows
the subspace allowed by $H_{0}$ value from Nevalainen \& Roos (1998) \label{omega_plot}}

\clearpage

\plotone{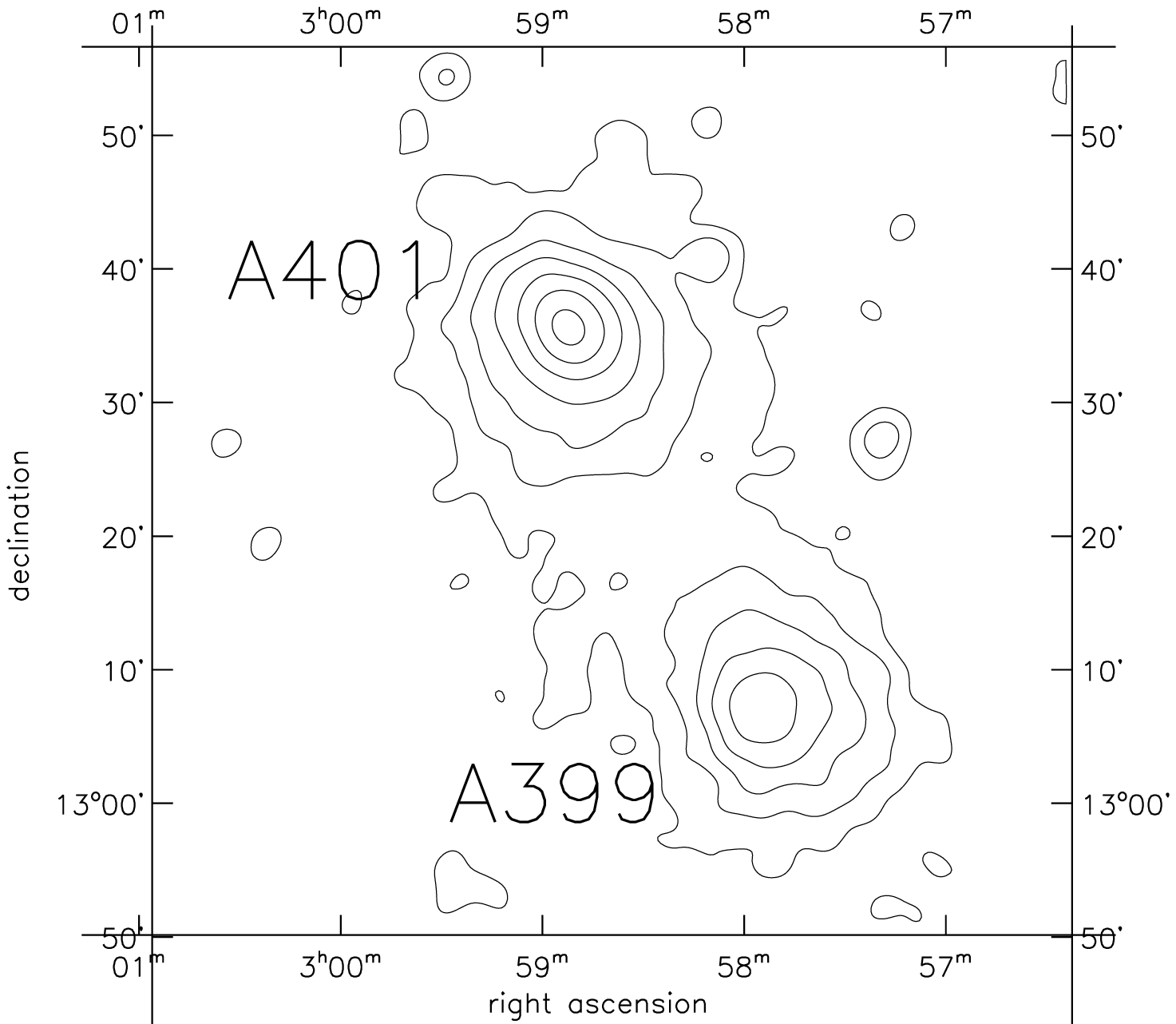}
\clearpage

\plotone{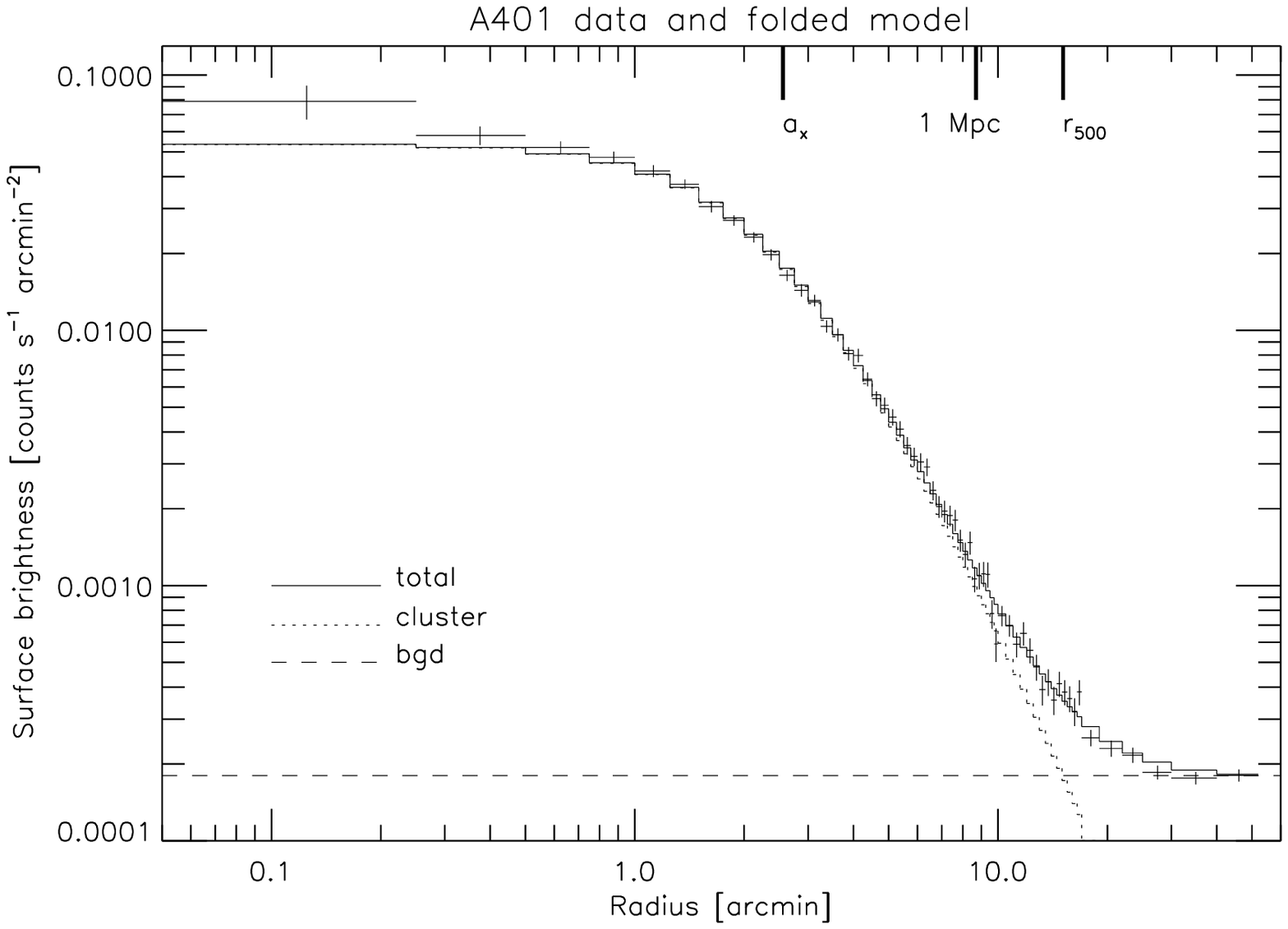}
\clearpage

\plotone{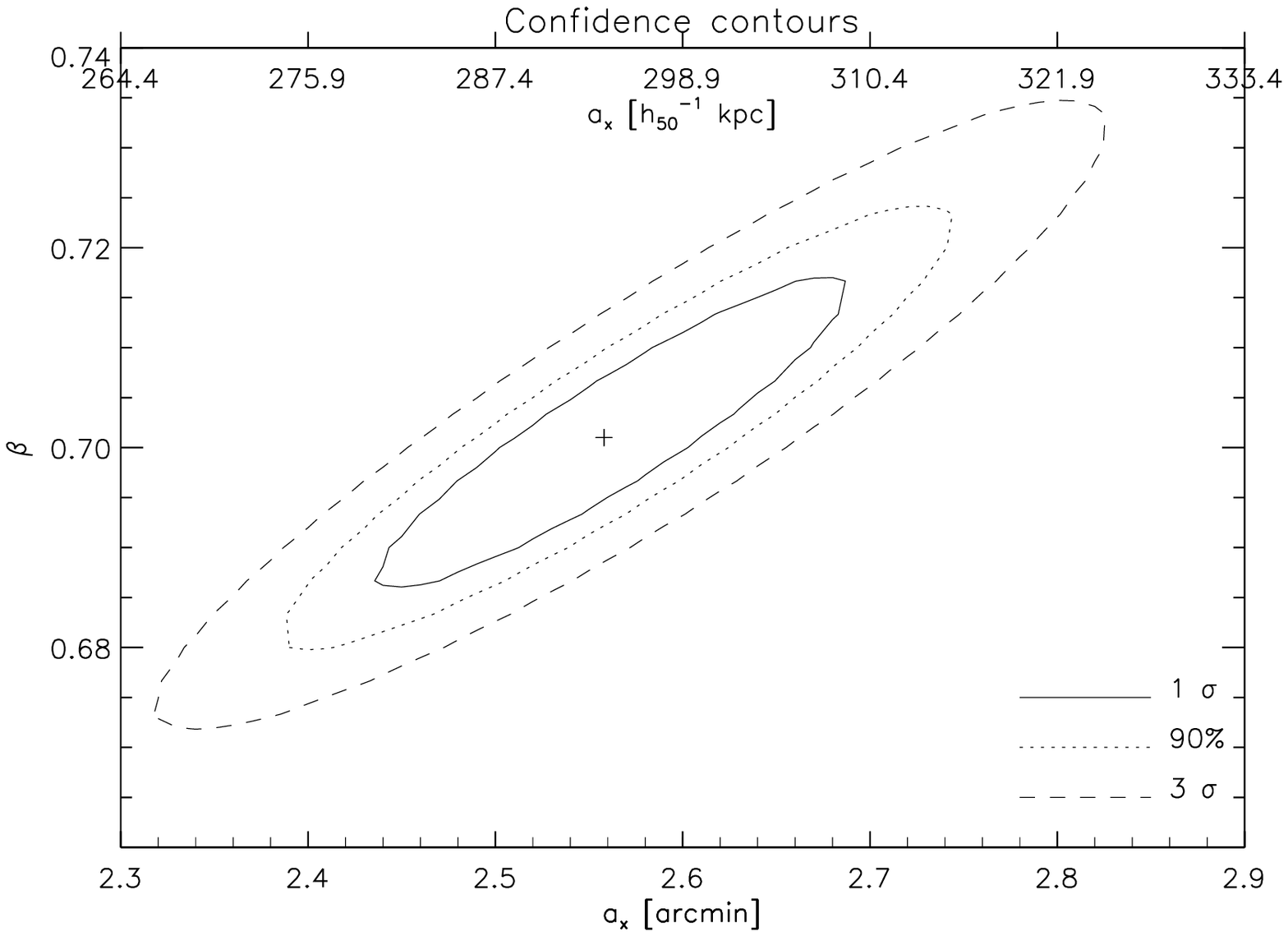} 
\clearpage

\plotone{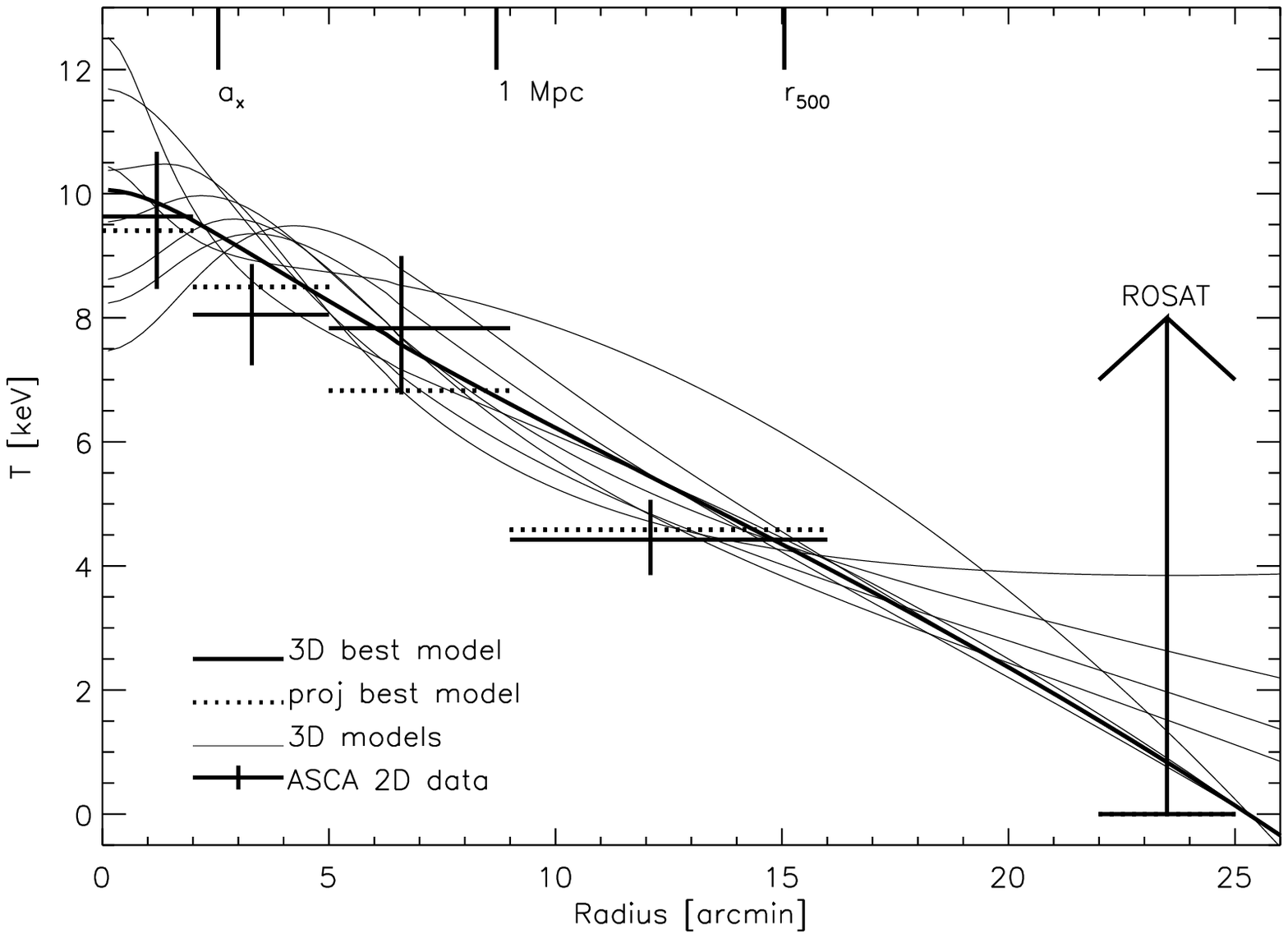}
\clearpage

\plotone{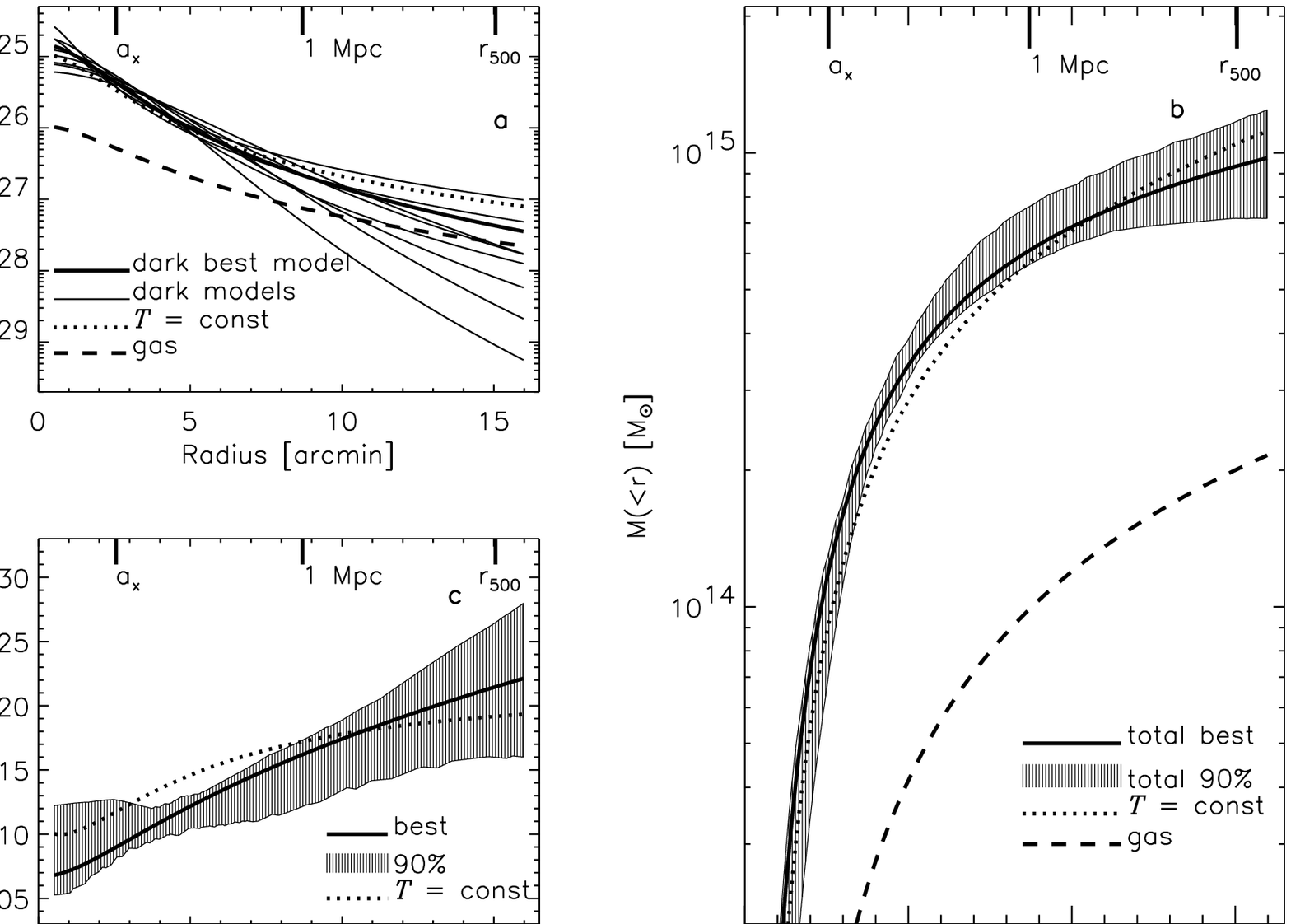}
\clearpage

\plotone{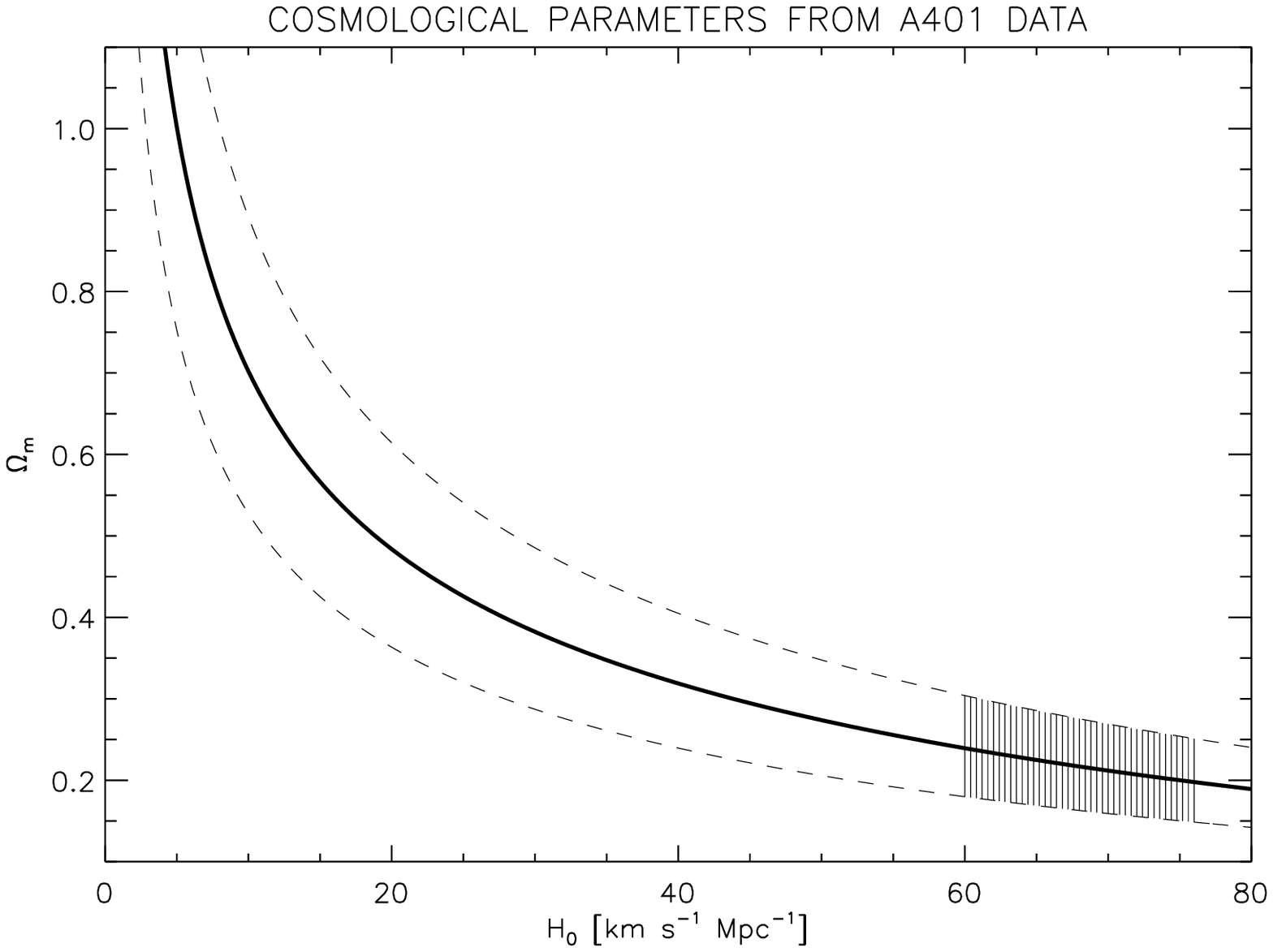}
\clearpage

\end{document}